\documentclass[fleqn,usenatbib]{mnras}

\usepackage{newtxtext,newtxmath}

\usepackage[T1]{fontenc}
\usepackage{orcidlink}
\DeclareRobustCommand{\VAN}[3]{#2}
\let\VANthebibliography\thebibliography
\def\thebibliography{\DeclareRobustCommand{\VAN}[3]{##3}\VANthebibliography}

\usepackage{graphicx}	
\usepackage{amsmath}	
\usepackage{soul}
\usepackage[
singlelinecheck=false 
]{caption}

\defcitealias{CY2016}{CY16}

\title[B-field strength using DCF with GSA]{Magnetic field measurement from the Davis-Chandrasekhar-Fermi method employed with Atomic Alignment}

\author[Pavaskar et al.]{
Parth Pavaskar $^{\orcidlink{0000-0003-3400-191X}}$,$^{1,2}$
Huirong Yan $^{\orcidlink{0000-0003-2560-8066}}$,$^{1,2}$\thanks{E-mail: huirong.yan@desy.de}
Jungyeon Cho $^{3}$
\\
$^{1}$Deutsches Elektronen-Synchrotron DESY, Platanenallee 6, 15738 Zeuthen, Germany\\
$^{2}$Institut f{\"u}r Physik und Astronomie, Universit{\"a}t Potsdam, Karl-Liebknecht-Str. 24-25, Haus 28, 14476 Potsdam, Germany\\
$^{3}$Department of Astronomy and Space Science, Chungnam National University, Daejeon, Korea
}

\date{Accepted XXX. Received YYY; in original form ZZZ}

\pubyear{2023}

\begin{document}
\label{firstpage}
\pagerange{\pageref{firstpage}--\pageref{lastpage}}
\maketitle

\begin{abstract}
 The Davis-Chandrasekhar-Fermi (DCF) method is widely employed to estimate the mean magnetic field strength in astrophysical plasmas. In this study, we present a numerical investigation using the DCF method in conjunction with a promising new diagnostic tool for studying magnetic fields: the polarization of spectral lines resulting from the atomic alignment effect. We obtain synthetic spectro-polarimetry observations from 3D magnetohydrodynamic (MHD) turbulence simulations and estimate the mean magnetic field projected onto the plane of the sky using the DCF method with Ground-State-Alignment (GSA) polarization maps and a modification to account for the driving scale of turbulence. We also compare the method to the classical DCF approach using dust polarization observations. Our observations indicate that the modified DCF method correctly estimates the plane-of-sky projected magnetic field strengths for sub-Alfv\'enic turbulence using a newly proposed correction factor of $\xi'\in 0.35-0.75$. We find that the field strengths are accurately obtained for all magnetic field inclination and azimuth angles. We also observe a minimum threshold for the mean magnetic field inclination angle with respect to the line of sight, $\theta_B\sim 16^\circ$, for the method. The magnetic field dispersion traced by the polarization from the spectral lines is comparable in accuracy to dust polarization, while mitigating some of the uncertainties associated with dust observations. The measurements of the DCF observables from the same atomic/ionic line targets ensure the same origin for the magnetic field and velocity fluctuations and offer a possibility of tracing the 3D direction of the magnetic field.

\end{abstract}




\section{Introduction} \label{sec:intro}

The interstellar medium (ISM) has been extensively studied in the past due to its importance in a wide range of astrophysical phenomena. One particularly crucial aspect of the ISM is the interstellar magnetic fields, which significantly influence the dynamics of the plasma. In addition, the magnetic fields impact several processes, including but not limited to plasma turbulence \citep[][]{GS95,Cho2000, CL03}, star formation \citep[][]{Crutcher2012, McKee07, Fissel2016}, stellar feedback, cosmic-ray transport and acceleration \citep[][]{Schlickeiser02_book, YL2002,YL2004}, accretion disk dynamics, astrophysical jets, and the chemical evolution of the galaxy \citep[see, e.g.][]{GHY16}. Therefore, accurately measuring the interstellar magnetic fields and their contributions to these processes is crucial in developing consistent theories. However, this measurement is not trivial.

On length scales shorter than the coherence scale of interstellar magnetic fields, the total field can be decomposed into two components; the (global) mean field with a preferential direction and the (local) turbulent field. While there are methods that utilize polarization information to measure the magnetic fields, e.g., the Davis-Chandrasekhar-Fermi method \citep[][hereinafter the DCF method]{Davis1951,CF1953} and the Polarization-Intensity gradient method \citep{Koch2012}, their probes typically rely on the polarization of emission/absorption arising from magnetically aligned dust. Although widely accepted as the conventional polarization diagnostic, dust alignment measurements may not be completely accurate owing to a number of uncertainties and inconsistencies \citep[see, e.g.][]{Reissl2014}. For instance, an obvious caveat with the conventional DCF method is the utilization of measurements of the line-of-sight (LOS) velocity and polarization from separate targets, i.e., the Doppler shift of spectral lines and the polarization of aligned dust emission or absorption. While modifications have been made to the DCF method to improve its accuracy \citep[][]{Heitsch2001, Falceta2008, Hildebrand2009, Houde2009, CY2016, Federrath2016,Skalidis2021}, this inconsistency is typically not addressed in the studies, making it necessary for other methods to be developed to be used in complement with the current techniques to trace the magnetic fields.

Several past studies have shown that in the presence of anisotropic optical pumping, the alignment of angular momenta of atoms and ions in the plasma can lead to the polarization of atomic spectral lines \citep{YL2006,YL2007,YL2008,YL2012,ShangguanYan2013, Zhang22018}. The UV or optical pumping by an anisotropic radiation field can cause uneven population distribution on the ground/metastable states and align the angular momenta of the atoms. In the presence of an external uniform magnetic field, the atoms are realigned owing to the fast magnetic precession. The resultant spectral lines from the aligned states are thus polarized toward the magnetic field. This effect, named atomic alignment or Ground-State-Alignment (GSA), is a powerful diagnostic in the study of the magnetic fields in the ISM. Both 3D direction and tomography can be retrieved by GSA \citep{YL2012, Yan2019}. More recently, polarized absorption lines from thr ground-state have been identified in a Post-AGB binary system 89Her, giving the observational confirmation of the applicability of the GSA effect \citep{Zhang2020}.

In this paper, we present a study that utilizes polarization observations and line width measurements from the same spectral lines to measure the magnetic field strength. We employ 3D simulations of magnetohydrodynamic (MHD) turbulence to obtain synthetic polarization observations arising from the Ground-State-Alignment (GSA) effect, which we then combine with the Davis-Chandrasekhar-Fermi (DCF) method to estimate the plane-of-sky (POS) projected magnetic field strength. We compare our new technique to the traditional DCF method, including cases of non-perfectly aligned polarized spectral lines.

Our work is organized as follows: we provide a brief explanation of the DCF method and the GSA effect in \S{2} and \S{3}, respectively. In \S{4}, we describe the simulation setup and the numerical methods used in this study. In \S{5}, we present our observations and results. Finally, we summarize our work in \S{6}.

\section{The modified DCF method}

The DCF method \citep{Davis1951, CF1953} is one of the most commonly used techniques for measuring the magnetic fields in a wide range of astrophysical systems, including molecular clouds, HII regions, and the interstellar medium in general. The method is based on the assumption that the magnetic field in a given region is in a state of equipartition with the turbulent motion of the gas inside it \citep{CF1953}. According to the method, the strength of the mean magnetic field projected onto the POS is given by:

\begin{equation}
\label{eq:dcf}
B_{0,\mathrm{pos}}=\xi \sqrt{4\pi\bar{\rho}}\,\frac{\delta \mathrm{v}_{\mathrm{los}}}{\delta\phi}
\end{equation}

\noindent where $\bar{\rho}$ is the mean density, $\delta v_{\mathrm{los}}$ is the velocity dispersion along the LOS and $\delta\phi$ is the dispersion in the angle between the turbulent and the mean magnetic fields projected on the POS. This angle dispersion is typically measured as the dispersion in the observed polarization vectors, while the LOS velocity fluctuations can be measured from the widths of optically thin emission lines. The constant $\xi$ is a correction factor usually taken to be $\sim 0.5$ \citep{Heitsch2001, Ostriker2001} or lower \citep{Liu2021}. The expression is derived from the condition that in Alfv\'enic turbulence, there exists an equipartition between the kinetic and magnetic energy densities, i.e. the root-mean-square (rms) fluctuations of the velocity and the magnetic field are related. In addition to Alfv\'enic (incompressible) turbulence, the DCF method also assumes that the velocity and magnetic field fluctuations are isotropic, and that the turbulent magnetic field energy is much smaller than the global B-field energy.

The POS-projected observables are always LOS-integrated, resulting in intrinsic limitations of the observed signals due to LOS averaging effects \citep{Zweibel1990, Myers1991}. This effect depends on the number of individual turbulent eddies along the LOS. The error is typically seen as an exaggerated alignment or ordering of polarization vectors, meaning that the polarization angles do not give accurate approximations of $\phi$. The error usually leads to an underestimation of $\delta\phi$ or an overestimation of the measured field strength $B_{0,\mathrm{pos}}$ in the DCF method.

\citet{CY2016} (hereafter \citetalias{CY2016}) found that this overestimation is roughly equal to a factor of $\sqrt{\mathrm{N}}\approx \sqrt{\mathrm{L_{los}/L_f}}$, where N is the number of independent eddies along the LOS, $\mathrm{L_{los}}$ is the length of the system along the LOS, and $\mathrm{L_{f}}$ is the driving scale of turbulence. They proposed a modified DCF method to account for the averaging effects, given by:

\begin{equation}
\label{eq:dcf_mod}
B_{0,\mathrm{pos}}=\xi' \sqrt{4\pi\bar{\rho}}\,\frac{\delta V_{\mathrm{c}}}{\delta\phi}
\end{equation}

\noindent where $V_{\mathrm{c}}$ is the normalized velocity centroid of the optically thin line, 
and the modified correction factor $\xi' \sim 0.7 - 1.0$. The velocity centroid is defined at the $i^{\mathrm{th}}$ LOS by

\begin{equation}
\label{eq:centroid}
\left. V_{\mathrm{c,i}}= \frac{\int v_{\mathrm{los}} \, I_i(v_{\mathrm{los}}) \, dv_{\mathrm{los}} \, }{ \int I_i(v_{\mathrm{los}}) \, dv_{\mathrm{los}}} \right. .\\
\end{equation}

\noindent $I_i$ is the optically thin emission line profile for the LOS. Since it has been shown by \citetalias{CY2016} that the modification makes the DCF method invariant to the turbulence driving scale, we shall use the modified method (equation~\ref{eq:dcf_mod}) for all the following numerical tests involving the DCF technique in this paper.

\section{Ground State Alignment} \label{sec:GSA}

\begin{figure}
    \centering
    \includegraphics[width=8.5cm]{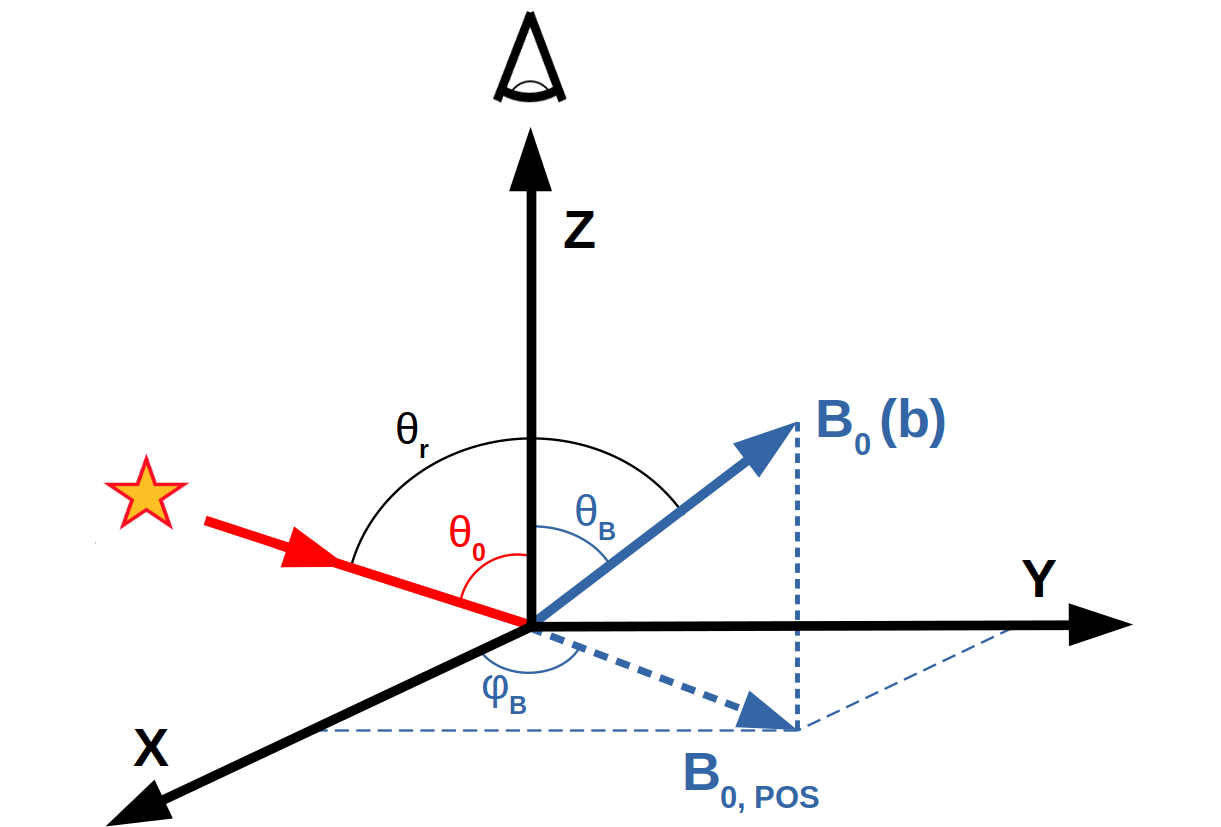}
    \caption{The geometry of our numerical setup. B$_0$ represents the mean magnetic field, and the LOS is fixed in the Z direction, with the X-Y plane representing the POS. The star symbol represents the source of anisotropic radiation, which is considered to be coming from an infinite distance and is parallel. The angle between the magnetic field and the LOS is denoted by $\theta_B$, while the angle between the projection of the magnetic field on the POS and the X axis is denoted by $\phi_B$. The angle $\theta_0$ follows the same logic for the radiation field direction with respect to the LOS. The angle between the magnetic field and radiation field is denoted by $\theta_r$.}
    \label{fig:geometry}
\end{figure}

In a typical ISM region where radiation sources, such as massive stars, are embedded in the diffuse plasma, atoms and ions in the plasma are continuously excited through optical pumping. When radiation excitation dominates, the occupation of the atoms/ions is determined by the optical pumping rate. In the case of anisotropic radiation, the net angular momentum in the photons is transferred to the atoms. If the collisional excitation rate is significantly lower than the radiative excitation rate, the angular momentum transfer causes the atoms to align along the direction of the incident radiation at the rate of the radiative pumping. Furthermore, if the Larmor frequency is larger than the radiative pumping rate in the presence of an external magnetic field, the atoms will be realigned due to fast magnetic precession. This condition can realistically be fulfilled in the diffuse ISM. For micro-Gauss scale magnetic fields in the diffuse medium, the atoms can only be aligned in their ground and/or metastable states. The magnetic realignment (parallel or perpendicular to the B field) depends on the angle between the mean magnetic field and the radiation field direction, $\theta_r$, and the resulting degree of polarization also varies with the magnetic field inclination, $\theta_B$ (the angle between the magnetic field and the LOS). As a result, information on the direction of the magnetic field is encoded in the polarization arising from the aligned atoms and ions. In the case of absorption from the atoms aligned in their ground or metastable states, the polarization direction directly traces the magnetic field in the plane of the sky \citep{YL2006,YL2012}. For atoms with fine structures, sub-millimeter fine-structure transitions are also polarized in the same manner \citep{YL2008}. \footnote{See review by \citet{YL2012} for the list of absorption, emission as well as fine structure lines and their maximum polarization fractions.}.

For a background unpolarized pumping source, the GSA effect will only produce linearly polarized lines. The degree of polarization for transitions from $J_1$ to $J_2$ for both absorption and fine structure emission lines is given by \citet{YL2006,YL2008}

\begin{equation}
\label{eq:gsa}  
P=\frac{1.5 \, \sigma_0^2(J_1, \theta_r) \, \mathrm{sin}^2\theta_B \, \omega^2_{J_1 J_2}}{\sqrt{2}+\sigma_0^2(J_1, \theta_r)\,(1-1.5\,\mathrm{sin}^2\theta_B)\,\omega^2_{J_1 J_2}}
\end{equation}

\noindent where $\theta_r$ and $\theta_B$ are the polar coordinates of the magnetic field vector (see Fig.~\ref{fig:geometry}). The alignment parameter $\sigma_0^2 \equiv \rho^2_0 / \rho^0_0$, is the normalized dipole component of the ground state density matrix, where $\rho^{2,0}_0$ are the irreducible density matrices. The parameter $\omega^2_{J_1 J_u} \equiv \{1,1,2;J_1,J_1,J_u\} / \{1,1,0;J_1,J_1,J_u\}$ is determined by the atomic structure \citep[see][]{YL2012}. The sign of $\sigma_0^2$ determines the orientation of the polarization vector with respect to the magnetic field. A positive polarization degree means a parallel orientation, while a negative polarization degree indicates a perpendicular orientation. This sign change or flipping of the polarization vector orientation happens at a specific $\theta_r=54.7^\circ , \, (180 - 54.7)^\circ$, also known as the Van Vleck angle \citep{VVleck1925,House1974}. In real observations, this leads to the magnetic field being mapped with a $90^\circ$ degeneracy (VV degeneracy from here onward). In principle, this degeneracy can be broken if more than two lines are identified in the observations.

\section{Numerical method} \label{sec:numerics}

\begin{table}
	\centering
	\caption{ Descriptions of MHD simulation cubes. The Alfv\'en velocity is in code units i.e. in units of $1/\sqrt{4\pi \bar{\rho}}$. The Alfv\'en and sonic Mach numbers are given by $v/v_A$ and $v/c_s$, respectively, where $v$ is the rms velocity and $c_s$ is the sound speed.}
	\label{tab:cubed}
	\begin{tabular}{lcccc}
		\hline
		Name & Grid & Alfv\'en & Alfv\'en Mach & Sonic Mach \\
         & size & velocity ($v_A$) & number ($M_A$) & number ($M_s$)\\
		\hline
        d\_024 & $512^3$ & 0.24 & 0.80 & 1.68 \\
        d\_030 & $512^3$ & 0.30 & 0.66 & 1.98 \\
        d\_040 & $512^3$ & 0.40 & 0.50 & 2.00 \\
        d\_050 & $512^3$ & 0.50 & 0.40 & 2.00 \\
        d\_070 & $512^3$ & 0.70 & 0.26 & 1.82 \\
        d\_080 & $512^3$ & 0.80 & 0.20 & 1.65 \\
        \hline
	\end{tabular}
\end{table}

The numerical method used in the study is divided into two parts: the generation of synthetic polarization maps, and the analysis of the maps using the modified DCF method described in \S{2}. In this work, we calculated 3D MHD turbulence simulations with spatial grids of $512^3$ pixels. The set of sub-Alfv\'enic simulations ranging from $M_A=0.26$ to $M_A=0.8$ is generated using the high-order finite-difference PENCIL-code\footnote{\url{http://pencil-code.nordita.org}}. Turbulence is driven solenoidally with an isothermal equation of state i.e $P=\rho c_s^2$ where $\rho$ is the density and $c_s$ is the sound speed. The solenoidal (divergence-free) forcing ensures that the energy fraction of the incompressible Alfv\'en mode dominates over the compressible magnetosonic (fast and slow) modes in the turbulence. The details of all the simulations used in this work are given in Table~\ref{tab:cubed}.

The geometry of the numerical setup is shown in Fig.~\ref{fig:geometry}. We fix the LOS along the Z direction of the simulation box so that the POS is the X-Y plane. The incoming radiation is considered to be parallel and originating from an external source in the X-Z plane.

\begin{figure}
    \centering
    \includegraphics[width=\linewidth]{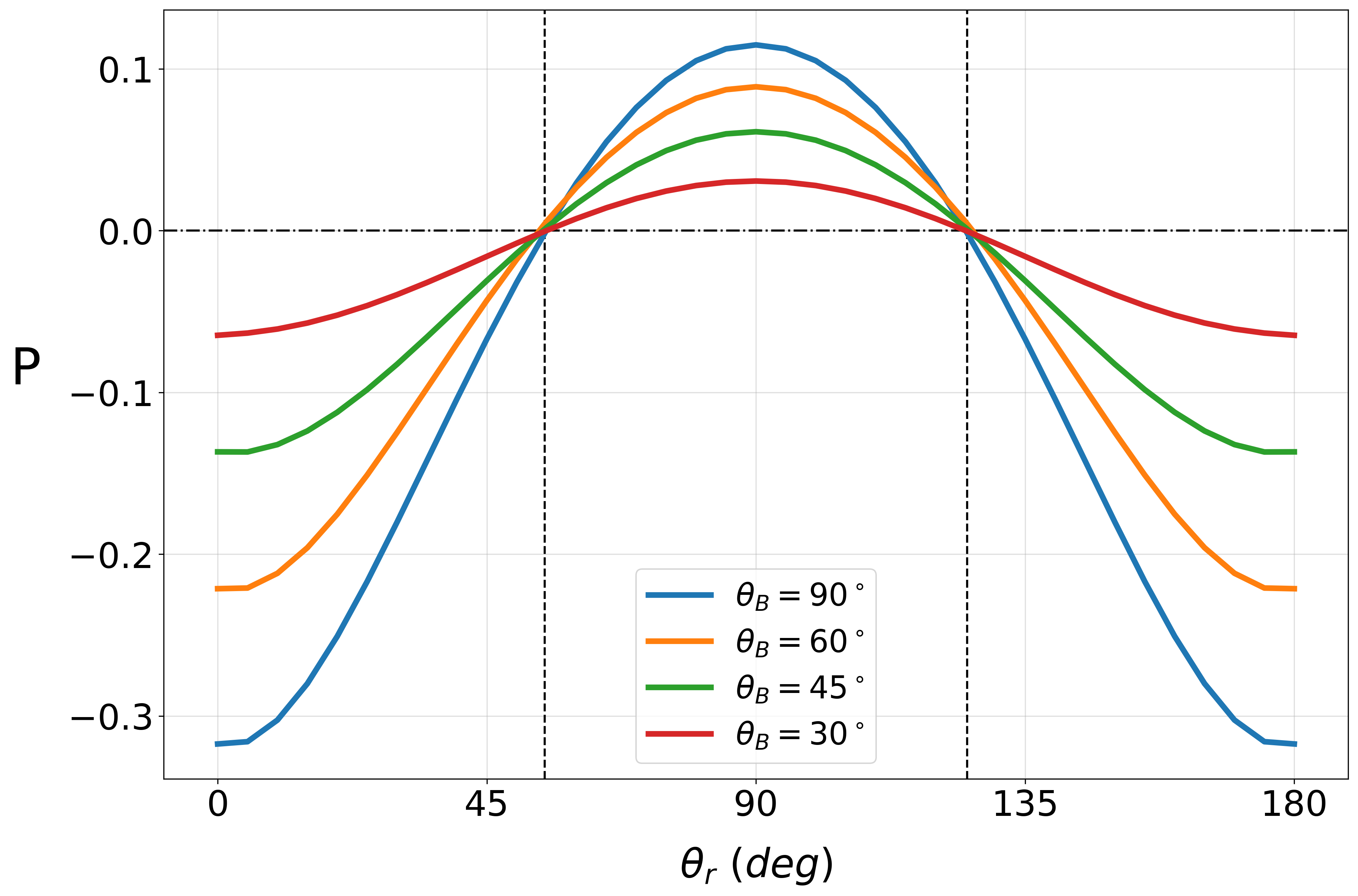}
    \vspace{8pt}
    \caption{This figure shows the computed degree of polarization versus $\theta_r$ for the fine structure emission line [C II]$\,\lambda157 \, \mu\mathrm{m}$. The colors represent different $\theta_B$. The positive and negative polarization fractions indicate parallel and perpendicular alignment to the magnetic field, respectively. The Van Vleck angles (54.7$^\circ$, (180-54.7)$^\circ$), at which the transition takes place are marked by vertical dotted lines.}
    \label{fig:gsa_c2_pol}
\end{figure}

\subsection{Calculation of synthetic Stokes maps} \label{sec:numerics_maps}

The line polarization to be simulated from the GSA effect depends on the directions of the radiation field  and the local magnetic field. Without loss of generality, we choose the [C II]$\,\lambda157 \, \mu\mathrm{m}$ (C$^+$) fine structure emission line for our synthetic observations. \citet{Zhang22018} have shown that C$^+$ can reach high maximum polarization (up to almost 30\%). Moreover, C$^+$ is commonly observed in the diffuse ISM. The degree of polarization arising from the GSA effect for the C$^+$ line for different mean field inclinations ($\theta_B$) is shown in Fig.~\ref{fig:gsa_c2_pol}. As is evident, the sign of the polarization fraction $P$ changes at the VV angle (shown by the vertical dotted lines), which means that the polarization vector is aligned parallel to the magnetic field direction in the range $\theta_r=(54.7^\circ,\, 125.3^\circ)$, and perpendicular for other inclinations. To get the synthetic polarization, we first obtain the $\theta_B$ and $\theta_r$ at each grid point relative to the local magnetic field direction, and calculate the total polarization degree $P(\theta_r, \theta_B)$ using the transition equation~(\ref{eq:gsa}). Next, we calculate the local Stokes parameters $q_z$ and $u_z$ at the grid points as follows
\begin{gather}
    q_z = P(\theta_r, \theta_B) \, \rho \,\, \mathrm{cos}\,2\phi_B \label {eq:local_q}\\
    u_z = P(\theta_r, \theta_B) \, \rho \,\, \mathrm{sin}\,2\phi_B \label {eq:local_u}
\end{gather}

\noindent where $\rho$ and $\phi_B$ are the local density and the local magnetic field azimuth angle, respectively. 

Typically in the previous studies of the modified DCF method \citep[\citetalias{CY2016}; ][]{YC2019}, the dust grains responsible for the polarized emission are assumed to be perfectly aligned with the external magnetic field. While this is done for the sake of simplicity in calculations, this assumption itself can cause errors in magnetic field measurement. Owing to the 90$^\circ$ VV degeneracy in the polarization of spectral lines by the GSA effect, such an assumption is not straightforward for our case. Consequently, we want to make sure that this assumption does not have a significant impact on the estimation of the magnetic field strength. For this reason, we generate two kinds of line polarization Stokes maps, one of the realistic scenario where the local Stokes parameters are weighted by the quantitative polarization fractions (first term on the right hand side of equations~(\ref{eq:local_q}) and (\ref{eq:local_u})) using equation~(\ref{eq:gsa}), and one with perfectly aligned atoms where the polarization fraction is assumed to be 1.

To simulate the LOS averaging in observations, we integrate the Stokes parameters along the LOS (which is the Z direction) to get the observed Stokes parameters

\begin{gather}
    Q = \int_z q_z  \, dz \label{eq:stokes2}\\
    U = \int_z u_z  \, dz \label{eq:stokes3}
\end{gather}

\noindent Since the background source is unpolarized, lines arising from the GSA effect will be linearly polarized i.e. the Stokes V $=0$. After the integration, we obtain 2D Stokes maps with the line averaged Stokes parameters Q and U. For the purpose of comparison with the conventional DCF approach utilizing dust polarization measurements, we also generate synthetic dust polarization maps following the method from \citet{Fiege2000} \citep[see also][]{Zweibel1996, Heitsch2001}.

While previous numerical studies involving the DCF method have tested the applicability of the technique in various systems like the ISM and star forming regions, the effect of the orientation of the mean magnetic field, and especially the LOS inclination, is usually neglected. For studies that use dust polarization as a measure of the local field dispersion, it is common practice to assume a mean field aligned with the POS \citep{Ostriker2001, Padoan2001}. While it is helpful to consider such a case to simplify calculations and calibrate the methods, it rarely reflects the real astrophysical environments. We examine all the possible geometries of the system in our study. This is achieved by generating a range of synthetic polarization maps while rotating the simulation box each time such that the mean magnetic field vector B$_0$ scans across the entire solid angle, covering all possible orientations, i.e. $\theta_B \in [0,\, \pi]$, $\phi_B \in [0,\, 2\pi]$, with respect to the radiation field. The rotation of the simulation box is performed using the Euler 3D rotation algorithm\footnote{\url{https://www.github.com/doraemonho/LazRotationDev}} similar to the one adopted in \citet{Lazarian2018} and \citet{Yuen2018}. Moreover, the polar angle of radiation field source $\theta_0$ is changed from 0 to $\pi/2$ in six equal steps to check the effect of the LOS inclination of the radiation field on the observed polarization signals. The range of polarization maps allow us to employ the DCF method using atomic-line polarization while studying the accuracy of the technique as a function of three distinct parameters, namely the Alfv\'en Mach number ($M_A$) , the magnetic field direction ($\theta_B$ and $\phi_B$) and the radiation direction ($\theta_0$).

\subsection{B-field estimation using DCF analysis} \label{sec:numerics_magf}

\begin{figure}
    \centering
    \includegraphics[width=\linewidth]{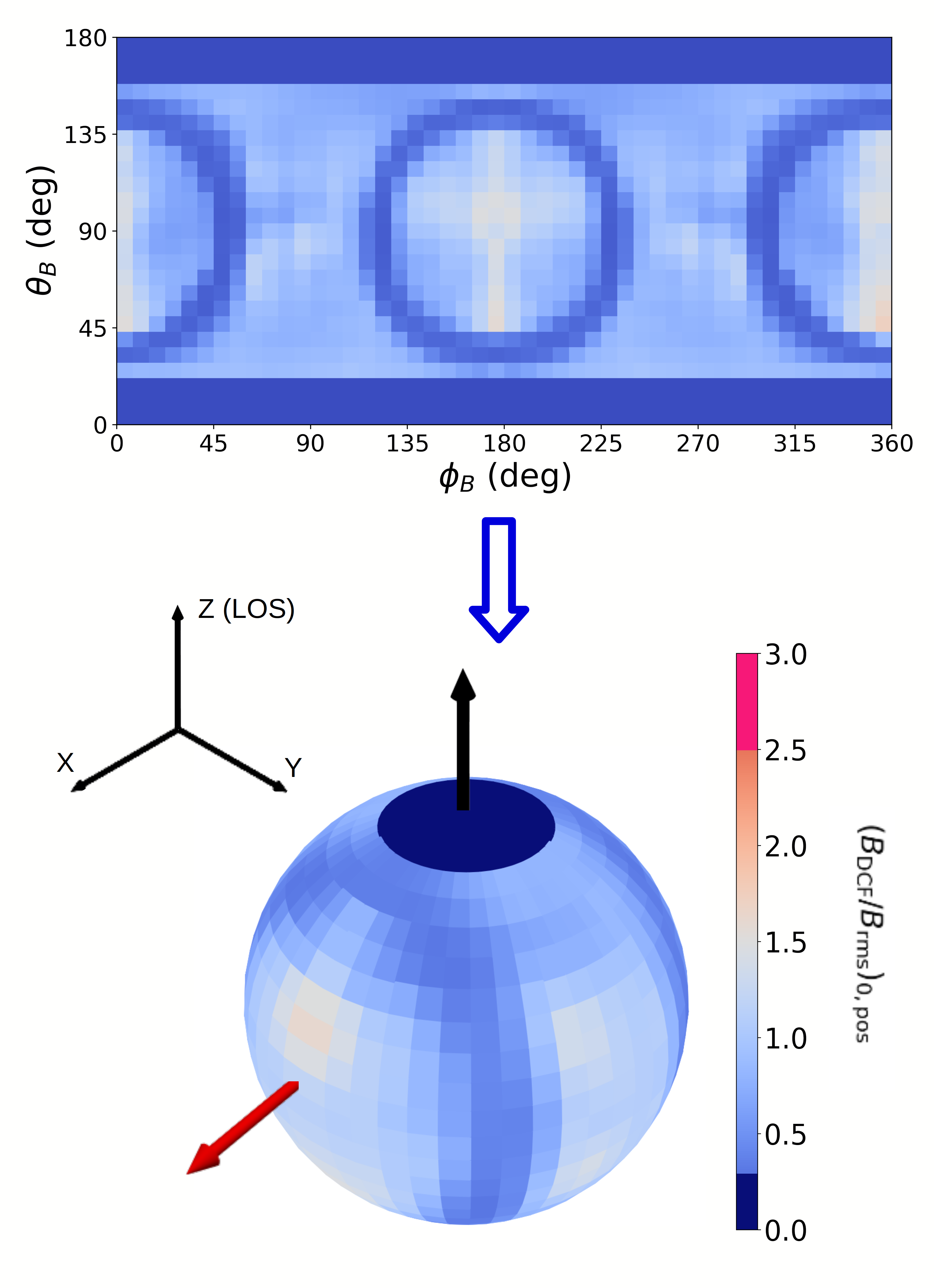}
    \vspace{8pt}
    \caption{An example of the rms normalized mean field strength measured using the modified DCF method at different orientations of the mean magnetic field vector. {\it Top}: Since the plot is in spherical coordinates, a 2D representation can be confusing and difficult to interpret. {\it Bottom}: We wrap the 2D heat map around a sphere such that every point on the sphere corresponds to a magnetic field strength measured when the mean magnetic field vector is pointing in that direction. The red and black arrows on the sphere represent the direction of the radiation field and the LOS, respectively.}
    \label{fig:plot_exp}
\end{figure}

With the 2D polarization maps, we can now apply the modified DCF method analysis to estimate the mean magnetic field strength using equation~(\ref{eq:dcf_mod}). Since the term $\sqrt{4\pi\bar{\rho}}$ is normalised to 1 in the MHD simulations, we require the two observables, i.e., the dispersion of the LOS velocity centroids $\delta V_c$ and the local magnetic field dispersion on the POS $\delta \phi$, to obtain the mean field strength. The centroids $V_c$ are calculated at each LOS using equation~(\ref{eq:centroid}) to obtain a velocity centroid "map". The dispersion is then calculated following the definition from \citetalias{CY2016} as follows

\begin{equation}
\label{eq:dvc}
\delta V_{\mathrm{c}}= \left( \frac{1}{n_{\mathrm{obs}}} \sum_{i=1}^{n_{\mathrm{obs}}} V_{\mathrm{c,i}}^2 - \left( \frac{1}{n_{\mathrm{obs}}} \sum_{i=1}^{n_{\mathrm{obs}}} V_{\mathrm{c,i}} \right)^2 \right)^{1/2} 
\end{equation}

\noindent where $n_{\mathrm{obs}}$ is the POS spatial resolution.

The POS local magnetic field dispersion can be estimated in the synthetic observations by measuring the deviation in the polarization vectors. The linear polarization fraction and angle can be recovered on each grid point (LOS) on the Stokes maps using
\begin{gather}
    P =  \frac{\sqrt{Q^2 + U^2}}{I} \label{eq:pol_frac} \\[2ex]
    \phi_P = \frac{1}{2} \, \mathrm{tan}^{-1}_2\frac{U}{Q}, \label{eq:pol_ang}
\end{gather}

 where $\tan^{-1}_2$ is the 2-argument arc-tangent function. To convert the Stokes maps into polarization maps, we transform $P$ and $\phi_P$ into polarization vectors for each LOS. We can then use the minimum variance method to estimate the mean polarization direction $P_0$. This is done to simulate real polarization observations, where the direction of the projected mean magnetic field is not necessarily known \footnote{while circular statistics are typically used to compute the dispersion from polarization maps \citep{Circular_st}, we notice that for sub-Alfv\'enic turbulence (with $M_A<1$), the difference between the angle dispersion calculated using  the circular standard deviation and the minimum variance method is negligible.}. The method involves computing the variance in the polarization vectors around arbitrary unit vectors in the range $(0^\circ, 180^\circ)$. The direction of the unit vector that has the minimum variance in the polarization vectors is chosen as the mean polarization direction, i.e

 \begin{equation}
P_0 = \mathop{\arg \min}\limits_{u\in (0, \pi)} \left( \sigma^2(P_i, u) \right)
\end{equation}

 \noindent where $P_i$ is the polarization vector for the i$^{th}$ line of sight, $u$ is an arbitrary vector in the range $(0, \pi)$, and $\sigma^2(P_i, u)$ is the variance of the polarization vectors around the vector $u$. The angle dispersion is then simply computed as $\delta \phi = \sigma (P_i, P_0)$. Finally, we can substitute these measures in equation~(\ref{eq:dcf_mod}) to obtain the POS projected mean magnetic field strength. Since we calculate the Stokes maps for all orientations of the mean magnetic field in the $\theta_B-\phi_B$ space, we can use the DCF analysis to estimate the B-field strengths as a function of $\theta_B$ and $\phi_b$ ($\theta_B$ and $\phi_B$ being the polar and azimuth angle, respectively). This is shown with a heat map in the top panel of Fig.~\ref{fig:plot_exp} with an example simulation. Since the plot is in spherical coordinates of the mean magnetic field vector, a 2D heat map does no represent the dependency accurately. We wrap the values around a sphere (bottom panel of Fig.~\ref{fig:plot_exp}) to make the plot more intuitive, such that the colorbar value at each point on the sphere shows the B-field strength estimated by the DCF method when the mean magnetic field vector in the system points in that direction. This approach provides a clearer representation of the effect of the B-field direction on the DCF method and helps to better convey the results. Thus, while the sphere itself does not mean anything in the geometry of the setup, it allows us to easily visualize the performance of the DCF method as a function of the mean magnetic field inclination and orientation. The line of sight (LOS) and the direction of the incoming parallel radiation is also depicted by black and red arrows, respectively. Consequently, the angle between the radiation arrow and an arrow pointing to any point on the sphere gives us the mean $\theta_r$ for the corresponding geometry.

\section{Results and discussion} \label{sec:results}

We divide the results section into three parts. Initially, we study the influence of the Alfv\'en Mach number on the estimated field strength using DCF with GSA polarization. In the second part, we examine how accurately the technique works in different magnetic field orientations. Lastly, we study the performance of our technique when varying the direction of the radiation source relative to the LOS.

\begin{figure}
    \centering
    \includegraphics[width=\linewidth]{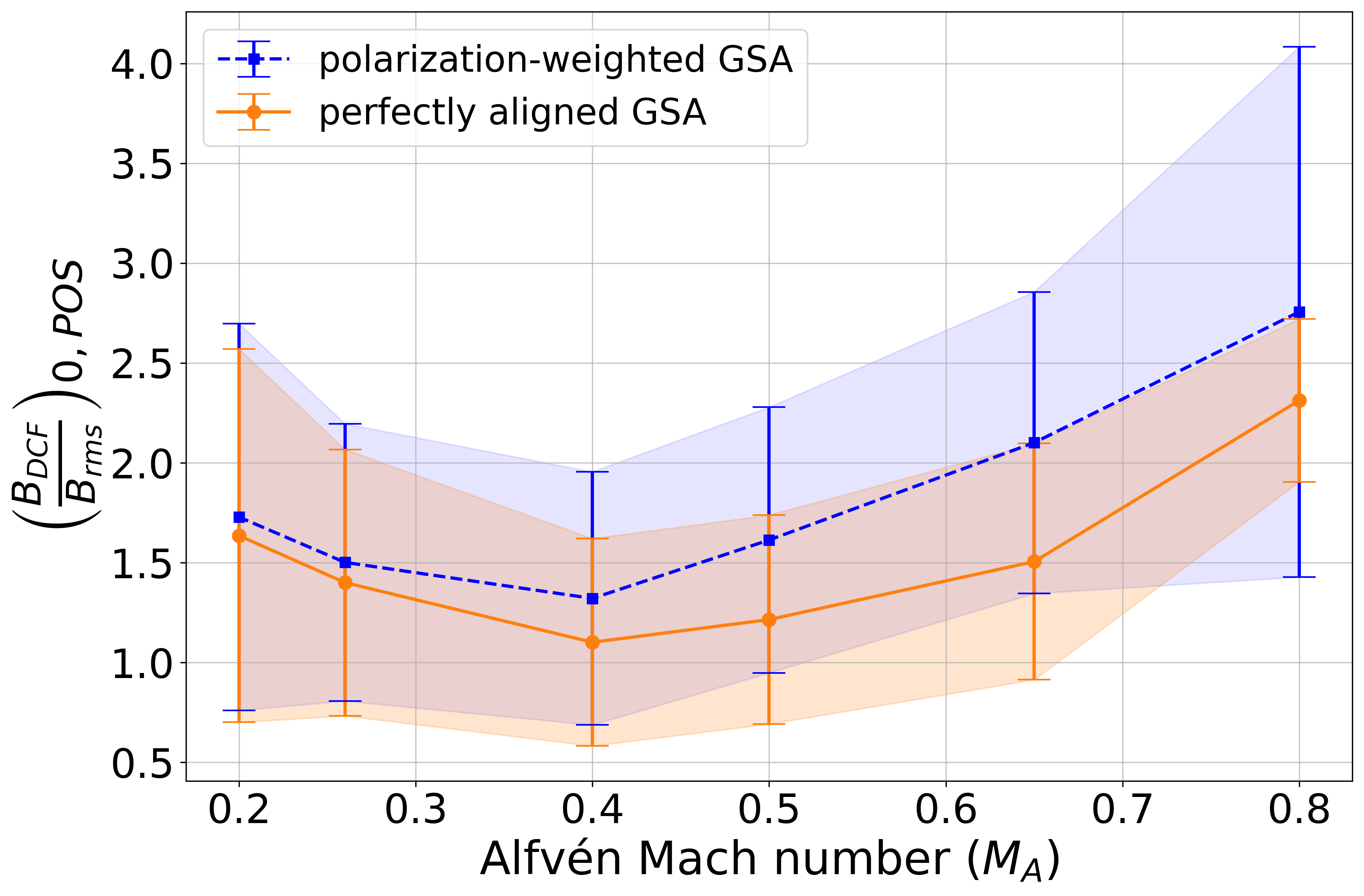}
    \vspace{8pt}
    \caption{The POS mean magnetic field strength for different Alfv\'en Mach numbers (data cubes) computed with (orange) and without (blue) assuming polarization from perfectly aligned atoms. The field strength is averaged over all values in the $\theta_B-\phi_B$ space, excluding the low inclination region at $\theta_B < 16^\circ$. The values are normalised with the POS projected rms magnetic field in the simulation box, such that a value of 1 represents the ideal measurement in the above plot. Radiation source direction is fixed at $\theta_0=90^\circ$ i.e. in the POS.}
    \label{fig:cbes_result}
\end{figure}

\subsection{Dependence on the Alfv\'en Mach number} \label{sec:results-ma}

\begin{figure*}
    \centering
    \includegraphics[width=0.9\linewidth]{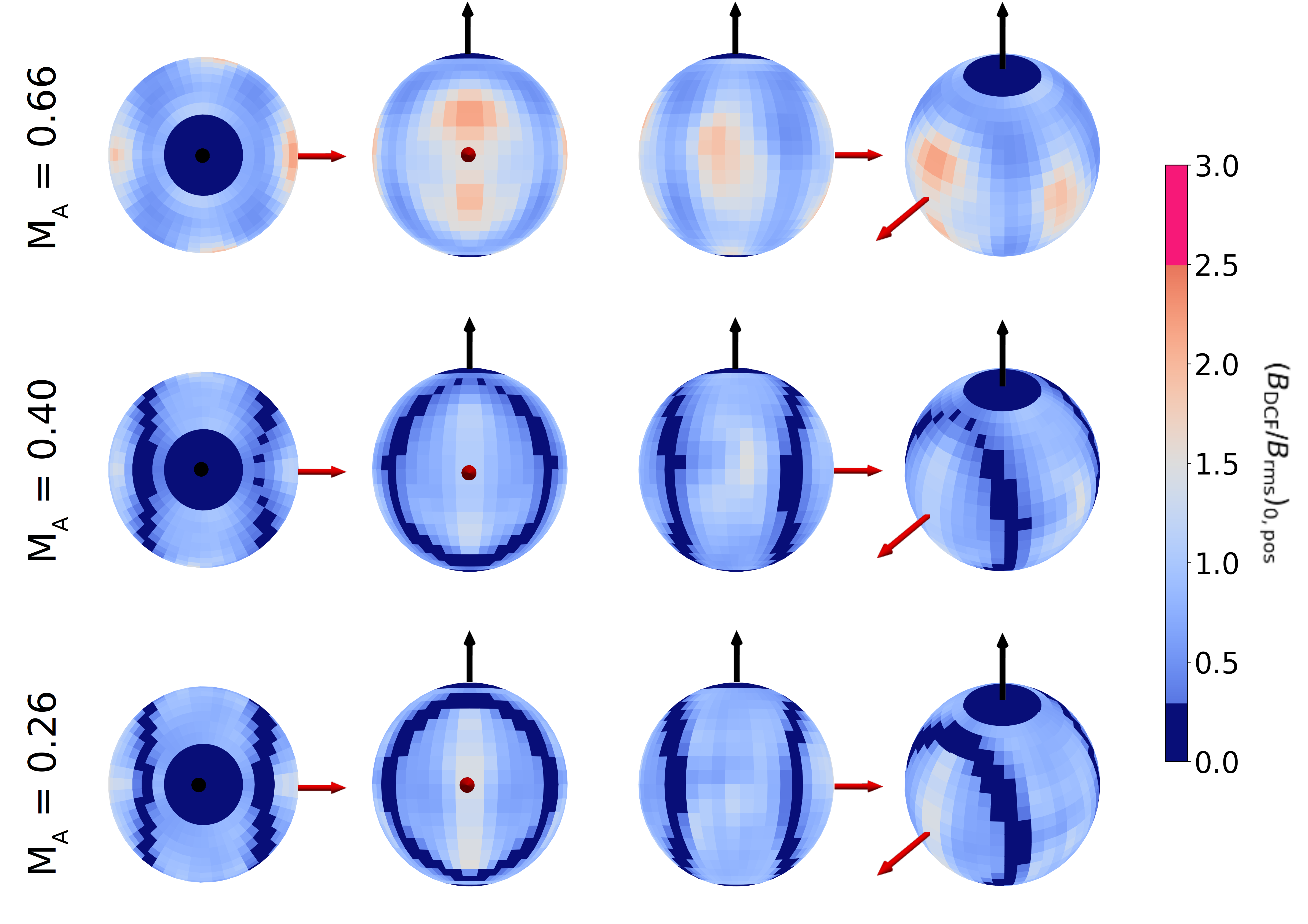}
    \vspace{8pt}
    \caption{The normalized POS mean magnetic field strength observed using the modified DCF method from polarization observations arising from GSA using a correction factor $\xi'=0.5$. The four columns are used to show the 3D distribution in $\theta_B-\phi_B$ space (which is done by rotating the numerical cube) in the X-Y, Y-Z, X-Z and isometric planes. The rows show three different Alfv\'en Mach numbers (from the top, $M_A=0.66, 0.40$ and $0.26$). The black and red arrows represent the LOS and the radiation field direction, respectively. }
    \label{fig:result_ma_3d}
\end{figure*}

In order to study the influence of the Alfv\'en Mach number, we consider the DCF estimated field strengths for all simulations, covering $M_A$ range in the sub-Alfv\'enic regime from $0.26$ to $0.8$. For this purpose, we average the B-field strengths over all possible mean B-field orientations. However, we choose to exclude all the cases where the magnetic field inclination with the LOS is smaller than a threshold angle $\theta_B < 16^\circ$ from the averages, since the DCF method has intrinsic limitation at small magnetic field inclination angles (see \S \ref{sec:results-mag} for further discussion). The radiation source is fixed in the POS i.e $\theta_0 = 90^\circ$. For comparison, we utilize both the perfectly aligned and realistic GSA polarization maps (see \S~\ref{sec:numerics_maps}) to measure the magnetic field dispersion in the DCF method. Lastly, we normalize the measured field strengths with the POS projected rms magnetic field strength from the simulations, which is the ideal value that we aim to measure. Fig~\ref{fig:cbes_result} shows the normalized field strengths measured using the modified DCF method utilizing the two types of synthetic GSA polarization observations (ideal and realistic alignment, see \S~\ref{sec:numerics_maps}) for all numerical simulations used in this study.

In both cases for perfect and realistic atomic alignment, the measured field strengths seem consistent for low $M_A$, with a rise in values as $M_A$ increases. Apart from a noticeable difference in the error bounds at high $M_A$, the values for perfect and non-perfect alignment are relatively similar in the sub-Alfv\'enic regime. Overall, the B-field predictions from the realistic GSA technique are slightly higher than the ideal counterpart for different $M_A$. This distribution can be explained through the intrinsic assumptions considered in the DCF method, which requires that the turbulent B-field energy is much smaller than the mean B-field energy. As $M_A$ increases in the turbulence, the turbulent field energy increases relative to the mean B-field energy. As a result, it is less likely that this assumption is satisfied. The highly turbulent plasma can also lead to abrupt changes in the mean field orientation along the LOS, which can contribute to high LOS averaging errors. From a general perspective, the method seems to typically overestimate the magnetic field strength by a factor of 1.3 to 2.5 in the sub-Alfv\'enic regime with similar error-bar spreads.

Following their modification to the DCF method, \citetalias{CY2016} proposed a correction factor $\xi' = 0.7 - 1$ based on the variation in their measured B-field strengths. However, the general trend in Fig~\ref{fig:cbes_result} indicates that the correction factor $\xi'$ in the modified DCF equation (\ref{eq:dcf_mod}) should be a function of $M_A$ when utilised with polarization from atomic alignment, rather than a constant correction as is typically considered for DCF using dust polarization. While determining the exact dependency of the $\xi'$ on $M_A$ for our technique requires further investigation into the method, it is apparent that the factor is lower than \citetalias{CY2016}. We propose a new correction factor $\xi' \in 0.35 - 0.75$ in case of sub-Alfv\'enic turbulence. In principle, our proposed correction factor is similar to the correction factor used in the conventional DCF method ($\xi$ in equation(\ref{eq:dcf}), typically taken $\sim0.5$). Although it should be noted that \citetalias{CY2016} only considered turbulence with $M_A \sim 0.6$ in their simulations, which could have influenced the resulting correction factor in their work.

\subsection{Influence of the magnetic field orientation on the DCF method} \label{sec:results-mag}

To investigate how the estimation from the DCF technique is affected by the mean magnetic field direction, we examine the B-field strengths for all magnetic field orientations in 3D (in the $\theta_B - \phi_B$ space). Fig.~\ref{fig:result_ma_3d} shows the B-field strength estimations from the DCF analysis with polarization from realistic GSA using a correction factor $\xi' = 0.5$ for the simulations with $M=0.66, \,0.4, \,0.26$, shown from top to bottom. The colorbar in Fig.~\ref{fig:result_ma_3d} shows the ratio between the DCF-measured B-field and the actual B-field after the POS projection. As described in section~\ref{sec:numerics_magf} (Fig.~\ref{fig:plot_exp}), the values at each point on the sphere show the magnetic field strength when the mean magnetic field vector points in that direction in the geometry given by Fig.~\ref{fig:geometry}, i.e. the measured field strength as a function of $\theta_B$ and $\phi_B$. $\theta_0$ is fixed at $90^\circ$ in all tests, and the LOS and the direction of the radiation field are shown using the black and red arrows, respectively.

As can be seen from the general distribution on the sphere, the observed mean field strengths are consistent with the rms values at most $\theta_B$ and $\phi_B$ with two noticeable exceptions: underestimations near the LOS and in ring-like regions around the radiation field direction and near the poles of the spheres (low $\theta_B$). The underestimation in the ring-like region is due the the VV effect, which is an intrinsic property of the atomic alignment process (see \S~\ref{sec:GSA}, and \citet{YL2006, YL2008} for a detailed description). When the angle between the mean magnetic field and the radiation direction is close to the VV angle, the fluctuations of the local magnetic field cause the sign of the GSA polarization fraction to change from point to point. This results in the local polarization vectors flipping abruptly between parallel and perpendicular alignment relative to the neighboring grid-points. A large number of such flips along the LOS can cause the magnetic field to be traced with significant inaccuracy after the LOS averaging. As a result, the dispersion in the polarization vectors is large, causing the B-field prediction from DCF to be underestimated. Since majority of the error from the VV effect arises due to the LOS integration of the polarization signal, it can be difficult to recognize the contribution of the VV effect in real observations if the geometry of the system is unknown. However, it is worth mentioning that the condition for the VV degeneracy, that the angle between the radiation field and the magnetic field $\theta_r \approx 54.7^\circ$, is a rare and special case of the system orientation that is unlikely to occur in majority of realistic astrophysical environments. Consequently, the phase volume of geometries where the observed polarization is affected by the VV degeneracy is limited. In principle, it is also possible to account for the VV effect by analysing the polarization signals in position-position-velocity (PPV) space and employing for e.g. a nearest neighbor filter to remove the LOS with large fluctuations in polarization vectors that do not correspond to large fluctuations in velocity. Such an approach, however, is not trivial and will be studied in detail in the future.

At small $\theta_B$, i.e. when the LOS and the mean magnetic field are close to being aligned, the projection of the mean field on the the POS is close to zero. Consequently, the polarization signals trace the turbulent field instead of the uniform field in the projection frame. Since the DCF method relies on the assumption that the dispersion in the polarization vectors is equal to the dispersion in the uniform field direction, this results in an intrinsic limitation of the DCF method at small inclination angles. In a detailed study and discussion regarding the inclination angle dependence in the DCF method, \citet{Falceta2008} observed that the DCF method heavily underestimates the field strength as $\theta_B$ approaches $0^\circ$ due to projection effects. More recently, \citet{LazYuen2020} proposed a modification to the DCF method to account for the inclination angle projection effect, given by

\begin{equation}
B_{0,\mathrm{pos}}=\sqrt{4\pi\bar{\rho}}\,\frac{\delta\mathrm{v}_{\mathrm{los}}}{\delta\phi} \, \frac{1}{\mathrm{sin}\,\gamma} \label{eq:LY20_eq}
\end{equation}

\noindent where $\gamma = \theta_B$ in this study. Essentially, the modification accounts for the difference between the strength of the projected magnetic field in the 2D plane and the total magnetic field strength through the correction factor ${\mathrm{sin}\,\gamma}$. We took the correction for the projection effect into account by normalising the values of magnetic field to the rms $B_{0,\mathrm{pos}}$ instead of $B_{0}$. It is worth mentioning that with the developments of new techniques which are capable of estimating the inclination angle of the mean magnetic field \citep[see, e.g.,][]{YYL23, Malik2023}, the mean field strength can be measured using equation~(\ref{eq:LY20_eq}). \citet{LazYuen2020} also showed that the method is only accurate at inclination angles larger than a minimum threshold, which they measured to be a function of $M_A$, given by $( 4\,\mathrm{tan}^{-1}\,(M_A/\sqrt{3}))$. Ensuring if the B-field inclination of the system is larger than this threshold condition can be particularly challenging with real observations, as it is notoriously difficult to estimate the $M_A$ of astrophysical plasma, even if the inclination angle is known. While we do see an $M_A$ dependence in our results in the form of increasing field strengths as we go to higher $M_A$, we find that it is not as strong as their measurement. Instead, we propose a minimum $\theta_B$ threshold independent of $M_A$ for all our simulations. Accordingly, we only consider orientations with $\theta_B > 16^\circ$ to make sure the projection effects do not influence the DCF estimate averages.

\subsection{Significance of the radiation field direction in GSA} \label{sec:results-rad}

\begin{figure*}
    \centering
    \includegraphics[width=0.9\linewidth]{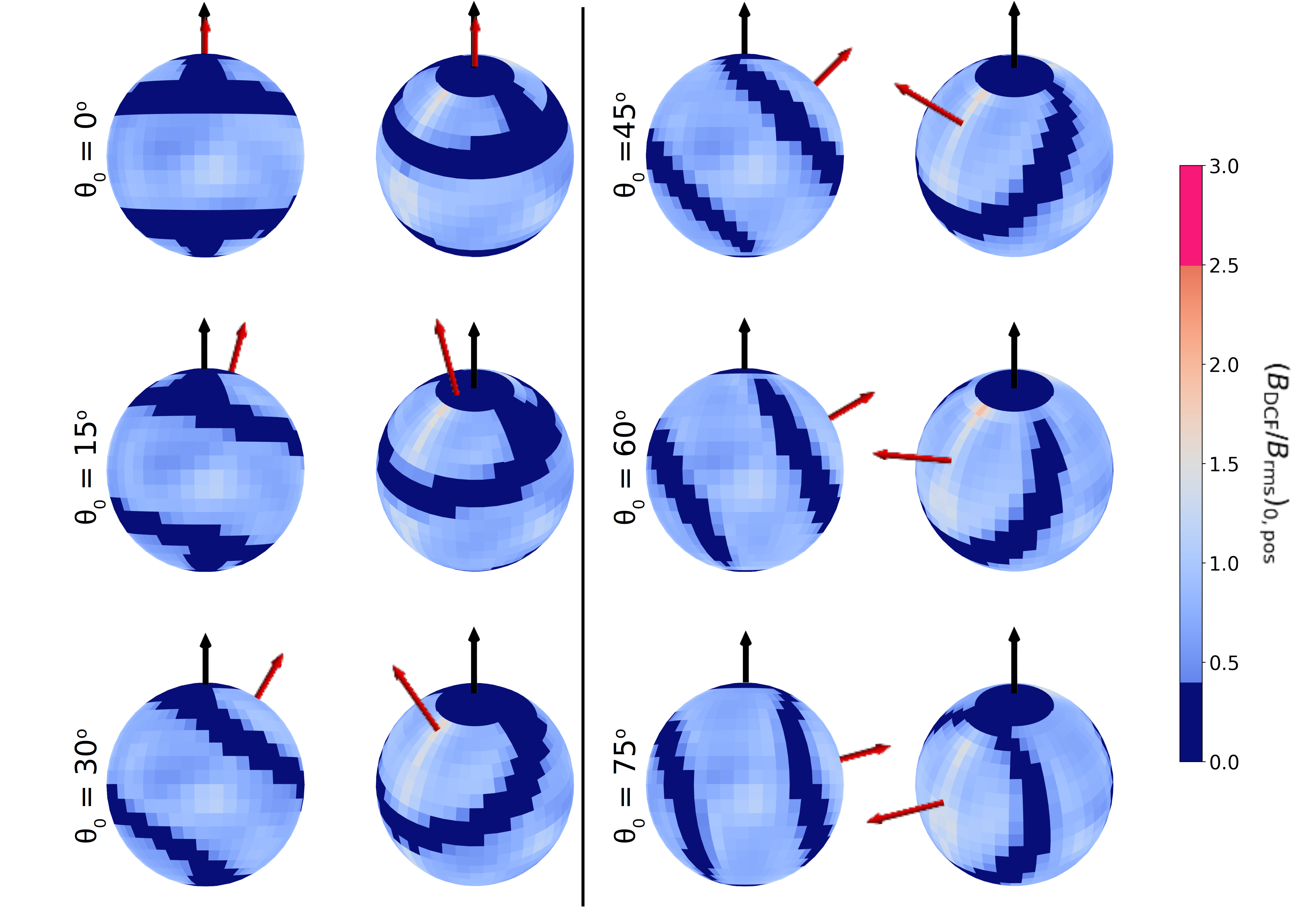}
    \vspace{8pt}
    \caption{ An example showing normalised field strengths measured in the $\theta_B-\phi_B$ space for different $\theta_0$ using the simulation d\_070 ($M_A=0.26$) and $\xi'=0.5$. For each $\theta_0$, the X-Z (first and third rows) and the isometric projections (second and fourth rows) are shown. The color bar normalization is similar to Fig.~\ref{sec:results-ma}. Red and black arrows depict the radiation field direction and the LOS, respectively.}
    \label{fig:result_theta0_3d}
\end{figure*}

The results presented in the previous section were limited to the special case of $\theta_0 = 90^\circ$, i.e. the external radiation source fixed in the POS. For the sake of completeness, we change the location of the radiation source in the X-Z plane and perform the DCF analysis on the generated GSA polarization synthetic maps to check the effect of the radiation field direction on the method.

Fig.~\ref{fig:result_theta0_3d}, which uses the simulation with $M_A=.26$ and a correction factor $\xi' = 0.5$, shows the distribution of field strengths in $\theta_B - \phi_B$ space for different $\theta_0$. It is evident that the change in $\theta_0$ changes the location of the VV regions, which is expected. However, the changing radiation source does not influence the DCF estimates at geometries in which the system is not affected by VV degeneracy (i.e. $\theta_r \neq $ VV angle). Although the method cannot resolve the magnetic fields in the VV region in its current state and requires some modifications, the VV orientation in itself is a special case geometry. Therefore, we expect the DCF method to work accurately with polarization from spectral lines regardless of the location of the radiation source, as long as the geometry of the system in real observations does not correspond to this special VV case. In addition, the correction factor of $\xi'\in 0.35-0.75$ discussed in section~\ref{sec:results-ma} applies to our method irrespective of the radiation source geometry.

\subsection{Comparison to dust polarization method} \label{sec:discussion-dust}

\begin{figure}
    \centering
    \includegraphics[width=\linewidth]{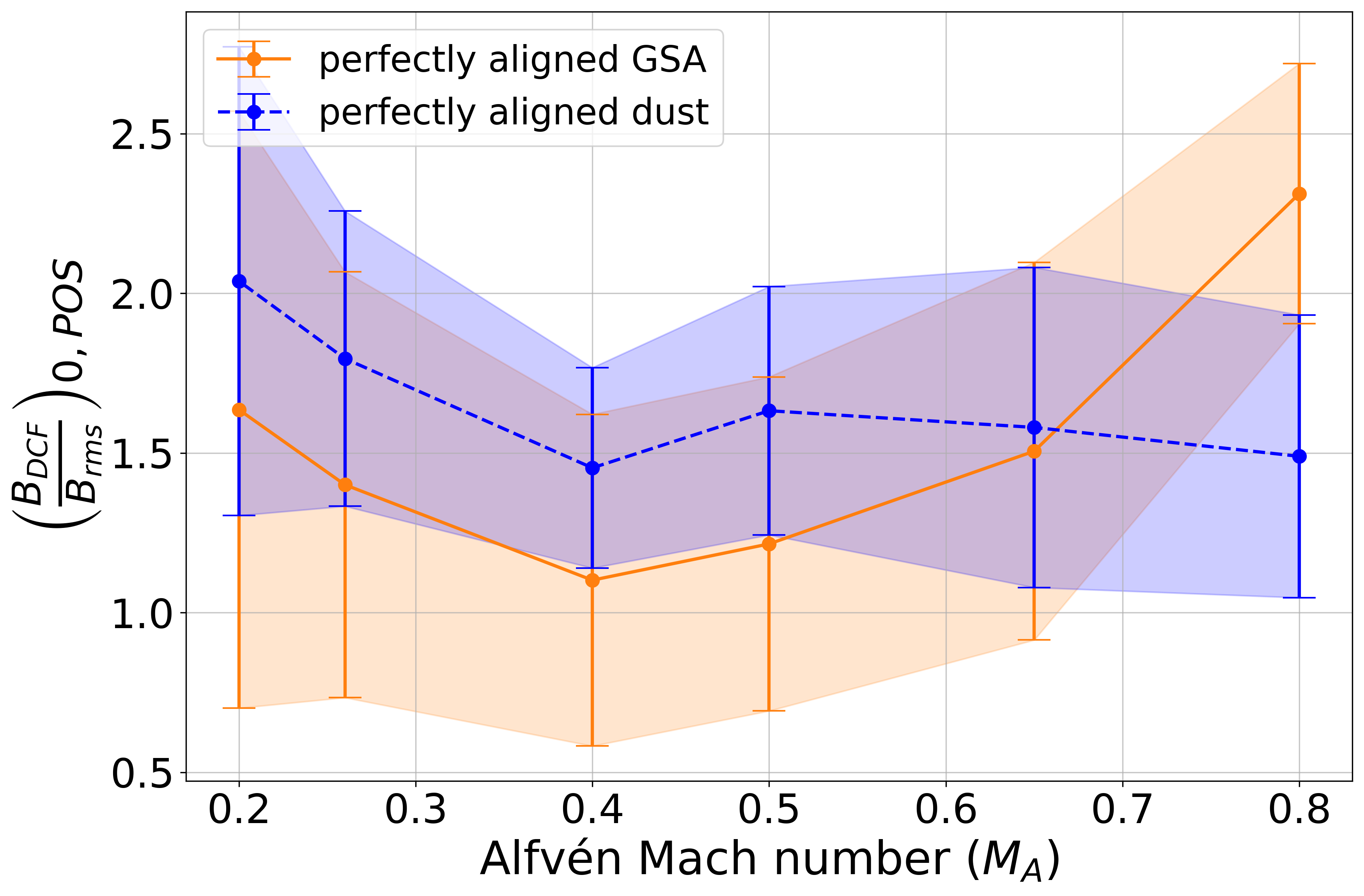}
    \vspace{8pt}
    \caption{Comparison between field strengths estimated by the DCF method utilizing polarization from perfectly aligned dust (blue) and GSA (orange). $\theta_0$ is fixed at 90$^\circ$ for the GSA measurements.}
    \label{fig:result_dst_comp_plot}
\end{figure}

\begin{figure*}
    \centering
    \includegraphics[width=0.9\linewidth]{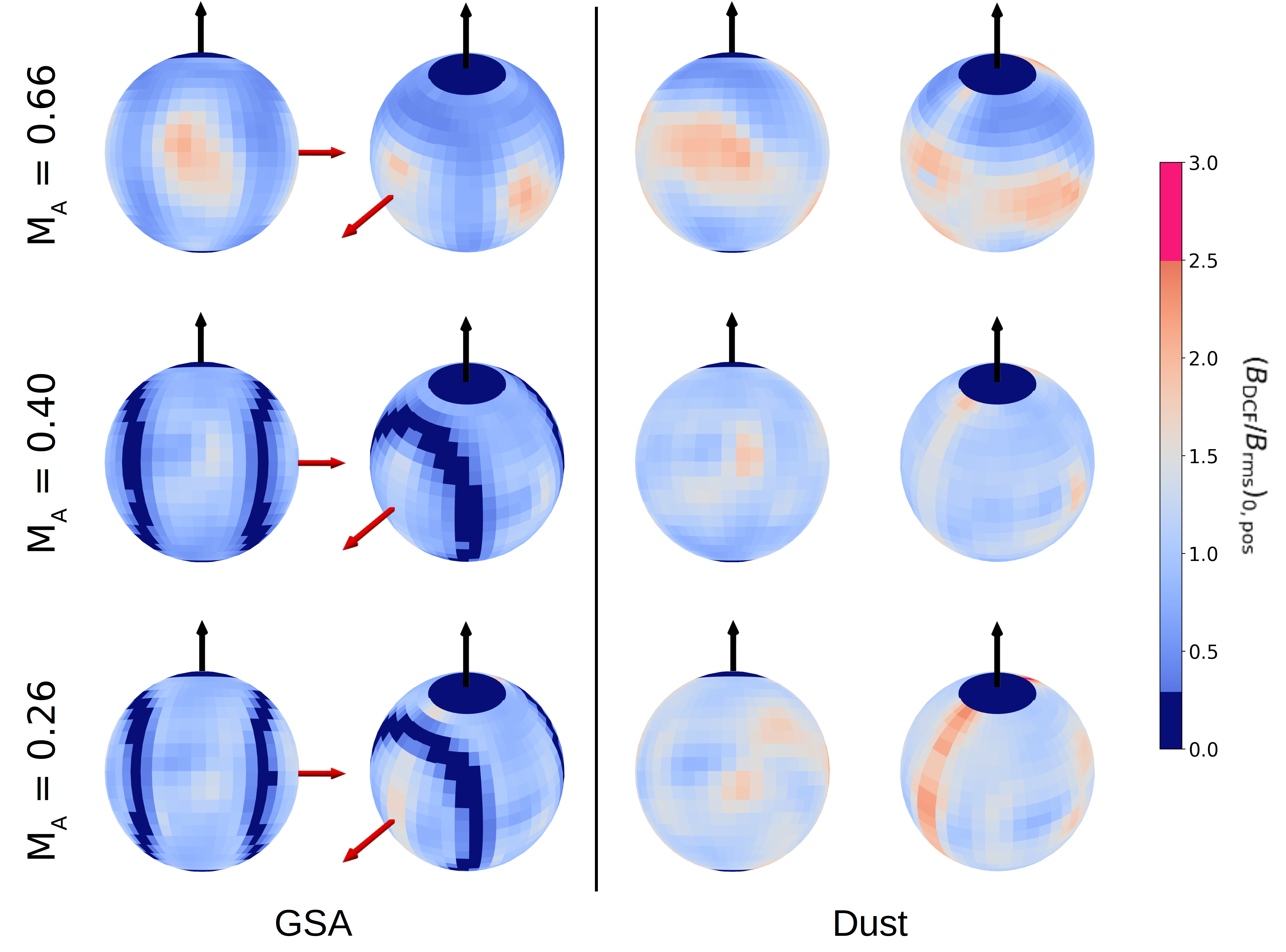}
    \vspace{8pt}
    \caption{A visual representation of the difference in field strengths measured at different $\theta_B$ and $\phi_B$ for DCF using GSA (left) and dust (right) polarization. A correction factor of $\xi'=0.5$ and $\xi'=0.8$ is used for GSA and , respectively following \citetalias{CY2016} and this work. The projection on the X-Z and isometric planes is shown. The normalization is similar to Fig.~\ref{sec:results-ma}. Red arrows depict the radiation field direction for GSA. Black arrows show the LOS.}
    \label{fig:result_dst_comp}
\end{figure*}

In most studies that employ the DCF method or its variations, both historically and presently, dust polarization observations are used to calculate the polarization angle dispersion. To examine how the DCF method utilizing polarization from atomic alignment compares to the classical dust approach, we simulated the synthetic polarization for both the mechanisms and use the DCF analysis to estimate the magnetic fields. The comparison is shown through the measured B-field strengths averaged over all magnetic field orientations versus $M_A$ in Fig.~\ref{fig:result_dst_comp_plot}. The averaging and normalization is performed similarly to Fig~\ref{fig:cbes_result} (see \S~\ref{sec:results-ma}). For the purpose of a balanced comparison, we consider only perfectly aligned atoms and dust grains. The comparison is also shown in the $\theta_B - \phi_B$ space for three different values of $M_A$ in Fig.~\ref{fig:result_dst_comp}, where the GSA method uses the correction factor $\xi'=0.5$ as calculated in this work (see \S~\ref{sec:results-ma}), while the dust alignment method uses $\xi' = 0.8$ as is suggested for dust polarization by \citetalias{CY2016}. 

From Fig.~\ref{fig:result_dst_comp_plot}, it is clear that in the sub-Alfv\'enic regime, the DCF technique using atomic GSA polarization measures the B-field strength with similar precision compared to the dust polarization method. The spread in errors also seem to be consistent for the two methods. From Fig.~\ref{fig:result_dst_comp}, it can be seen that the only difference in the two methods is the ring-like VV region for the GSA measurements, which is absent in the dust polarization. Although the VV regions can explain the lower averages for GSA in Fig.~\ref{fig:result_dst_comp_plot}, the estimations from the two approaches outside the VV regions appear to be highly consistent with each other. Despite the fact that dust polarization does not suffer from the VV degeneracy, it is important to note that the observations of dust maps are usually accompanied by their own uncertainties. Previous observations have shown that the dust grains are asymmetrical and align with the magnetic field lines along their shortest axis due to radiative torques \citep{Davis1949, Davis1951b,Cho2005, Lazarian2007, Andersson2015}, which means that realistic dust alignment traces the magnetic filed direction with a $90^\circ$ flip as well. Since the physical properties such as size and shape of the individual dust grains vary in the diffuse interstellar medium, the efficiency of the radiative alignment is different \citep[][]{DW1996,DW1997,LH07}. As a result, smaller grains which might not be aligned with the magnetic field, also contribute to the observed polarization signals. Especially in low density plasmas, the continuum dust polarization measurements can suffer from decreased signal-to-noise ratios due to low polarization fractions. In addition to physical properties, variation in the chemical composition of the dust also contributes to the unreliable measurements in the observations. 

Another challenge with dust polarization that can lead to inaccuracies, particularly in the DCF analysis, arises from the fact that optical/IR continuum observations are used for the polarization dispersion measurements, while velocity dispersion is obtained using optically thin spectral lines. While it is generally assumed that information from these sources originate in the same region in the magnetized medium, it may not always be true. It is straightforward that when the dispersions in velocity and polarization angles are calculated from two different layers along the LOS in the plasma, the DCF method does not give a correct estimate for the B-field strength at either of those layers. This particular uncertainty, as well as those arising from the diversity of the sizes, shapes and compositions of dust particles can be averted by using GSA observations, in which the same polarized atomic line can be used to gain information about the velocity and magnetic field fluctuations (polarization angle dispersion).

In addition, GSA could facilitate a new avenue in magnetic field diagnostics, namely the 3D tomography of the magnetic field in the ISM. In principle, this can be achieved by performing the DCF analysis using velocity slices, i.e, thin wavelength intervals or segments of the line profiles to get information about the magnetic field strength and orientation in the PPV space. However, this will require further numerical and observational studies, which we will address in the future.

\subsection{Testing the \citetalias{CY2016} method for atomic line polarization}

\begin{table}
	\centering
	\caption{Descriptions of MHD simulation cubes with different driving scale $L_f$. The driving wavenumber $k_f$ is in units of $ L_f / L$, where $L_f$ is the driving scale of turbulence and $L$ is the size of the simulation box along one axis.}
	\label{tab:k10}
	\begin{tabular}{l|cc}
		\hline
		Name & d\_040 & k\_024 \\
        \hline
        Resolution & $512^3$ & $512^3$ \\
        Alfv\'en velocity ($v_A$) & 0.40 & 0.12 \\
        Alfv\'en Mach number ($M_A$) & 0.50 & 0.50 \\
        Sonic Mach number ($M_s$) & 2.00 & 2.50 \\
        Driving wavenumber ($k_f$) & 2 & 10 \\
        \hline
	\end{tabular}
\end{table}

\begin{figure}
    \centering
    \includegraphics[width=\linewidth]{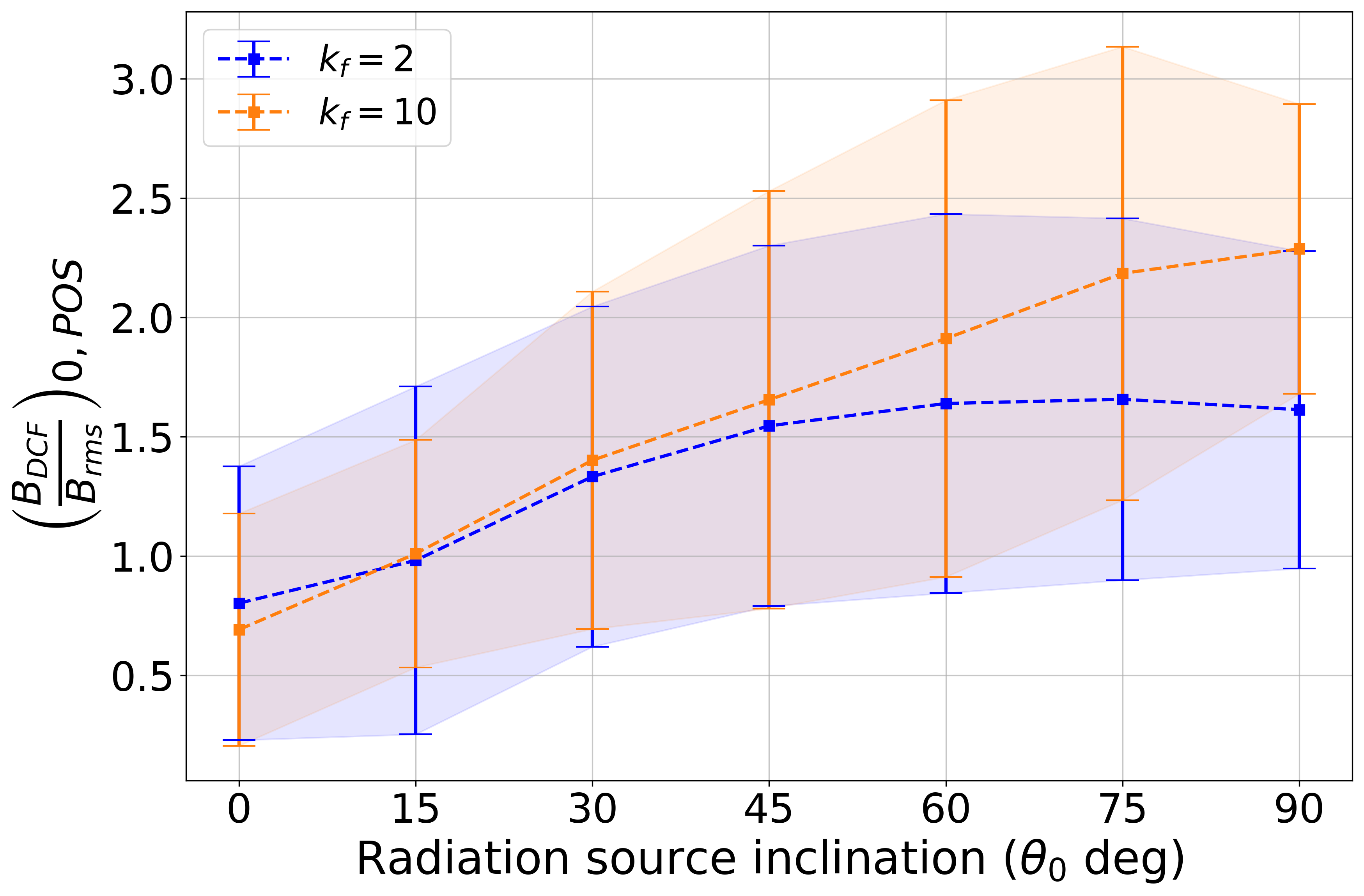}
    \vspace{8pt}
    \caption{The figure shows the normalised mean magnetic field strengths from DCF averaged over all magnetic field orientations versus the angle between the radiation source and the LOS ($\theta_0$). Simulation d\_040 with driving wavenumber $k_f=2$ is shown in blue, while k\_024 with $k_f=10$ is shown in orange.}
    \label{fig:result_k10}
\end{figure}

As a motivation for the modification to the DCF method, \citetalias{CY2016} showed that the DCF method is affected by the driving scale of the turbulence, and that the conventional DCF method overestimates the POS field strength by a factor proportional to the ratio of the LOS scale and the driving scale of turbulence ($\sim \sqrt{L_\mathrm{los}/L_f}$). We decided to check the efficiency of the modified DCF method while using atomic line polarization instead of dust polarization which was used in \citetalias{CY2016}. For this purpose, we utilized two separate simulations with similar $M_A$ and $M_s$, but different $L_f$. The details are given in Table~\ref{tab:k10}. Since the simulation box length is normalized to unity, the driving scale of turbulence ($1/k_f$) for the simulation d\_040 is larger by a factor of 5 than the simulation k\_024. According to \citetalias{CY2016}, a discrepancy of $\sim \sqrt{L_\mathrm{los}/L_f}$ in the conventional DCF method would lead to the POS mean field strength measured from k\_024 to be overestimated than that of d\_040 by a factor of $\sqrt{5} \approx 2.3$. We use the modified DCF method with line polarization from GSA to measure the POS field strengths for the two simulations for different magnetic field orientations, and plotted the averages over ($\theta_B, \phi_B$) against the $\theta_0$ in Fig~\ref{fig:result_k10}. The B-field strength estimations at low $\theta_0$ show no significant difference, while even as $\theta_0$ approaches $90^\circ$, the largest difference seen in the two simulations is by a factor of $\sim 1.4$. This is direct evidence that the modified DCF method from \citetalias{CY2016} corrects for the averaging effects from multiple eddies along the LOS, even when used with polarization from atomic alignment, and regardless of the radiation source orientation.

\section{Summary} \label{sec:sumary}

In this paper, we have employed the modified Davis-Chandrasekhar-Fermi method proposed by \citet{CY2016} along with synthetic polarization observations arising due to the Ground State Alignment effect \citep{YL2006, YL2008} in simulated magnetized plasma. Using 3D MHD turbulence simulations with varying plasma properties and geometries, we demonstrate the compatibility of the polarization observations of the GSA effect with conventional techniques like the DCF method and its variations. The method differs from the traditional DCF method by measuring the dispersion of the mean magnetic field direction through polarized spectral lines instead of continuum dust polarization measurements. The paper adopted the [C II] fine structure emission line without loss of generality. The method can be readily applied to archived and new spectro-polarimetry data covering wide wavelength ranges from UV to sub-millimeter \citep{YL2012}. 

The results from the numerical investigation of the method can be summarized as follows:

\begin{itemize}
    \item The modified DCF method using polarization maps from atomic ground state alignment gives consistently accurate estimates of the POS projected mean magnetic field strengths in the ISM. We propose a correction factor of $\xi'\in 0.35-0.75$ for sub-Alfve\'nic turbulence.

    \item The strength of the projected magnetic field in the plane of sky is obtained for all magnetic field inclination angles. We identify a minimum threshold angle for the magnetic field inclination with the line-of-sight of $\theta_B = 16^\circ$ below which the DCF method does not trace the magnetic fields accurately.

    \item The DCF method utilizing polarization measurements from atomic alignment is equally accurate as the conventional method utilizing dust polarization observations, while avoiding the uncertainties accompanied by dust alignment such as variations in grain size, shape and chemical composition. 

    \item The spectro-polarimetry combined with spectrometry from the same atomic/ionic lines not only improves the accuracy of the DCF method by ensuring the same origin for the magnetic field and velocity fluctuations, but can also potentially trace the 3D direction and strength of the local magnetic field.  

    \item The modified DCF method from \citet{CY2016} successfully accounts for the correction to the conventional DCF method due to the driving scale of turbulence irrespective of the polarization source. It is also invariant to the geometry of the local radiation source in case of atomic alignment by GSA.

\end{itemize}

In this study, we present a novel diagnostic for tracing the magnetic field fluctuations through atomic alignment, which can be used in conjunction with the DCF method to estimate the POS-projected mean magnetic field strength in the ISM. We demonstrate that our method improves the accuracy of the conventional DCF approach while taking into account the differences in atomic and dust polarization approaches.

\section*{Acknowledgements}

We would like to acknowledge the referee for the constructive suggestions. We acknowledge DESY (Zeuthen, Germany), a member of the Helmholtz Association HGF, and the  University of Potsdam for the support and the resources to make this work possible. We are grateful to Heshou Zhang and Bolu Feng for their contributions. We would also like to thank Ka Ho Yuen, Sunil Malik and Thiem Hoang for the helpful discussions.

\section*{Data Availability}

The data involved in this work will be shared upon reasonable request to the corresponding author.

\bibliography{sample631}{}
\bibliographystyle{mnras}

\end{document}